\documentclass[aps,pre,groupedaddress,twocolumn]{revtex4}
\pdfoutput=1

\newcommand{\deltat}{1}
\newcommand{\seltat}{1}

\usepackage{hyperref}
\usepackage{subfigure}
\usepackage{color}
\usepackage{graphicx}
\usepackage{natbib}
\usepackage{doi}

\begin{document}
\bibliographystyle{plainnat}

\expandafter\ifx\csname urlprefix\endcsname\relax\def\urlprefix{URL }\fi
\DeclareGraphicsExtensions{.pdf, .jpg}

\title{\large
 Hydrodynamics of the physical vacuum: dark matter is an illusion
}

\large
\author{Valeriy I. Sbitnev}
\email{valery.sbitnev@gmail.com}
\address{St. Petersburg B. P. Konstantinov Nuclear Physics Institute, NRC Kurchatov Institute, Gatchina, Leningrad district, 188350, Russia;\\
 Department of Electrical Engineering and Computer Sciences, University of California, Berkeley, Berkeley, CA 94720, USA
}


\date{\today}

\begin{abstract}
The relativistic hydrodynamical equations are being examined with the aim of extracting the quantum-mechanical equations (the relativistic  Klein-Gordon equation and the Schr\"odinger equation in the non-relativistic limit). In both cases it is required to get the quantum potential, which follows from pressure gradients within a superfluid vacuum medium. This special fluid, endowed with viscosity allows to describe emergence of the flat orbital speeds of spiral galaxies. The viscosity averaged on time vanishes, but its variance is different from zero. It is a function fluctuating about zero. Therefore the flattening is the result of the energy exchange of the torque with zero-point fluctuations of the physical vacuum on the ultra-low frequencies.
\\

\end{abstract}

\maketitle

\large

\section{\label{sec1}Introduction}

Richard Feynman in his lecture "the character of physical law"~\citep{Feynman1965} has outlined a series of crucial unsolved problems of modern physics. 
Among them there is an unfilled gap between quantum mechanics and the physics of large scale motion in the universe. Predominantly, it is a theory of quantum gravitation
The lecture encourages physicists to search for these solutions~\citep{Hajdukovic2013}.

Relativistic hydrodynamics~\citep{Font2008, LaigleEtAl2014, WilsonMathews2003} perhaps can cover this wide  region
of the scientific interest.
 Its scope of interests spreads from physics of plasma up to neutron stars, black holes, supernovae, compact binary systems, dense clusters, collapsing stars, the early universe, etc~\citep{WilsonMathews2003}.

Here we shall consider the relativistic hydrodynamical equations in the light of a description of the superfluid vacuum medium - the medium consisting of  enormous amount of virtual pairs of particles-antiparticles with opposite orientations of spins. That is, the pairs possess zero spin~\citep{Volovik2003}. 

We shall try to trace the emergence of the quantum equations (the Klein-Gordon equation and in the non-relativistic limit the Schr\"odinger equation) from the relativistic hydrodynamical equation + the continuity equation. It turns out, that for solving this task, we need to find the quantum potential through applying the internal pressure to Fick's laws. 

Second, we shall try applying the relativistic hydrodynamical equation for the description of the motion of spiral galaxies. Here we introduce a viscosity of the medium which is equal to zero in averaged on time but has a non-zero dispersion~\citep{Sbitnev2015b, Sbitnev2015c}. Due to this trick we find that the orbital velocity of the spiral galaxy attains flat level at large distances from the galactic center. Essentially, fluctuations of the viscosity about zero describes exchanging energy of the orbital rotation with the zero-point fluctuations of the physical vacuum on the ultra-low frequencies.
Figuratively speaking, the dark matter appears to us as the ultra-low vacuum fluctuations.

Here we first outline the general formulas which will then be used further.

The relativistic hydrodynamics starts from a description of the energy-momentum tensor~\citep{RezzollaZanotti2013}
\begin{equation}
  T^{\,\mu\nu} = (\epsilon + p){\mathit u}^{\mu}{\mathit u}^{\nu} + \,p \eta^{\,\mu\nu}.
\label{eq=1}
\end{equation} 
 Here $\epsilon$ and $p$ are the functions per unit volume.  
They are expressed in units of pressure, Joule/m$^3$~=~Pa.
The metric tensor $\eta^{\,\mu\nu}$ has the spacelike signature $(- + + +)$.
  The 4-velocity ${\mathit u}^{\mu}=\gamma(1,{\vec{\mathit v}})$ meets the condition
  ${\mathit u}_{\mu}{\mathit u}^{\mu}=-1$ where $\gamma = (1-{\mathit v}^2)^{-1/2}$~\citep{Romatschke2010a}.
 Here we follow the convention that the units of space and time are chosen such that $c=1$. However, sometimes we shall use the International System of Units where the speed of light has the customary value equal about $3\cdot10^{8}$ m/s.

The four-gradient (both covariant and contravariant) and the four-laplacian (the d'Alembert operator)
 compactly written read:
\begin{eqnarray}
\nonumber
&&\hspace{-8pt}
  \partial_{\mu} = \biggl( {{}{}} {{\partial\;}\over{\partial\,\tau}}, \nabla\biggr), ~
  \partial^{\,\mu} = \eta^{\,\mu\nu}\partial_{\nu} =  \biggl(- {{}{}} {{\partial\;}\over{\partial\,\tau}}, \nabla\biggr), \\
&&\hspace{-8pt}
 \partial_{\mu}\partial^{\,\mu} = \partial_{\mu}\eta^{\,\mu\nu}\partial_{\nu} =
 -{{}{}} {{\partial^2\;}\over{\partial\,\tau^2}} + \nabla^2.
\label{eq=2}
\end{eqnarray}
 Since  we adopted $c=1$  length is measured here in  unit of $\tau=tc$.
We shall write either $\partial\tau$ or $c\,\partial\,t$
when the customary units of time are needed.
In the last case we write ${\widetilde\partial}_{\mu}$ and~${\widetilde\partial}^{\mu}$.

An initial step in studying the relativistic hydrodynamical flows begins from writting the conservation laws of  energy and momentum $\partial_{\mu}T^{\,\mu\nu}=0$~\citep{WilsonMathews2003, LandauLifshitz1987}. Here we need to be careful. Our task is to write the relativistic equations of hydrodynamics, which would took into consideration quantum effects.  Appearance of the quantum potential in the course of computations gives guaranty that the computations are correct~\citep{Sbitnev2015b, Sbitnev2015c}.
This potential is a fundamental element, which creates a geometrodynamic picture of the quantum world in the non-relativistic domain, a relativistic curved space-time background, and the quantum gravity domain~\citep{Fiscaletti2012}.
 
 A final aim of the computations is to derive the  Klein-Gordon equation and the Schr\"odinger equation in the non-relativistic limit.
With this aim in  mind, we rewrite the conservation law in the following view
\begin{equation}
\partial_{\mu}(T^{\mu\nu}/\rho)=0.
\label{eq=3}
\end{equation}
Observe that the terms ${{(\epsilon+p)}/{\!\rho}}$ and  $p/\!\rho$ have dimensions of energy.
Here $\rho$ is the density distribution of the virtual particles populating the medium.
It is equal to number of the sub-particles per unit volume, $\rho = n/\Delta V$~\citep{JackiwEtAl2004}.

The article is organized by the following manner. 
Sec.~\ref{sect2} deals with the perfect fluid. It  consists of two subsections.
 In the first subsection we define relativistic  Fick's laws,
 and from here we compute the relativistic quantum potential. 
The second subsection considers the scalar field $S + \hbar\omega\gamma^{-1}t$
based on which the Klein-Gordon equation in the relativistic limit
and the Schr\"odinger equation in the non-relativistic limit are derived.
Sec.~\ref{sect3} considers a relativistic non-perfect fluid, that is the viscous fluid.
We get the relativistic Navier-Stokes equation and further we get the equation for the vorticity.
The vorticity and the orbital speed are considered for the case of the electron-positron pair.
Sec.~\ref{sect4} considers formation of the vortex on the galactic scales.
Formation of the flat orbital speed for the rotating spiral galaxy is the subject of consideration.
Sec.~\ref{sec5}  summarizes the results.

\section{\label{sect2}Relativistic perfect fluid:
 derivation of the Klein-Gordon and Schr\"odinger equations}

It turns out, that the equation~(\ref{eq=3}) can be reduced
 either to the Klein-Gordon equation or to the Schr\"odinger equation
 depending on the relativistic or non-relativistic limit.
However we first should show that the term ${\mathit p}/\rho$ is nothing like the quantum potential.
With this aim in  mind, we need to use  the relativistic Fick's laws.
Here we shall write the partial derivatives~(\ref{eq=2})
writing down $c\,\partial\,t$  rather than  $\partial\tau$.
Let us begin.

\subsection{\label{subsect2A}The Fick's law in the relativistic sector}

The first Fick's law in the relativistic limit has the following form
\begin{equation}
 {\mathit j}_{\mu} = - D{ {\widetilde\partial}_{\mu}}\rho_{_{M}}
\label{eq=4}
\end{equation}
 Here $D$ is the diffusion constant having the dimension of length$^2$/time. 
 The law~(\ref{eq=4}) says that diffusion flux vector, ${\mathit j}_{\mu} = ({\mathit j}_{0}, {\vec{\mathit j}})$, is induced by the inhomogeneous distribution of the mass density $\rho_{_{M}}$ within the 4D space-time volume. 
 Here 
$\rho_{_{M}}=M/{\delta V} = mn/{\Delta V}=m\rho$~\citep{JackiwEtAl2004, Sbitnev2015c}
and the mass $M=mn$ is that of the volume $\Delta V$ of the medium in question
($m$ is the mass of the single sub-particle
and $n$ is the amount of them populating the volume ${\Delta V}$~\citep{JackiwEtAl2004}).

Not to be confused the diffusion flux vector ${\mathit j}_{\mu}$ with the density current
${\mathit J}_{\mu} = \rho_{_{M}}{\mathit u}_{\mu}$~\citep{Font2008, RezzollaZanotti2013},
 where ${\mathit u}_{\mu}$  is the fluid 4-velocity and $\rho_{_{M}}$ is the proper rest-mass density.
 The diffusion flux vector ${\mathit j}_{\mu}$ can be seen as a result of scattering of the sub-particles
 on each other due to the collisions.

We need to define the diffusion coefficient~$D$.
First we introduce a hypothesis that all space-time is filled with virtual radiations with  frequency $\omega$.
So, we define 
\begin{equation}
  D = {{1}\over{\epsilon_{0}\mu_{0}\,\omega}} = {{c^{2}}\over{\,\omega}}.
\label{eq=5}
\end{equation}
Here $c=1/\sqrt{\epsilon_{0}\mu_{0}}$ is the light speed in the physical vacuum,
and $\epsilon_{0}$, $\mu_{0}$ are the absolute dielectric permittivity of classical vacuum and
the vacuum permeability, respectively.
 Observe that the zero-point energy of the radiation is $E =\hbar\omega/2$.
The radiation is due to motions of the virtual particles with the proper mass $m$. 
So we may write $E = mc^2$   
 and consequently
\begin{equation}
  D = {{\hbar}\over{2m}}  
\label{eq=6}
\end{equation}
 Here $\hbar$ is Planck's reduced constant.

In the light of this vision we may imagine that motions of the virtual particles have a weak dispersion around a certain  average. We may guess that the virtual particles on this level undergo collisions with each other like the Brownian particles. 
Here we draw on the hypothesis put forward by Edward Nelson 
about aether populated by the Brownian sub-particles~\citep{Nelson1966}. 
However, in contrast to the quasi-classical collisions of the Brownian sub-particles in our case the diffusion is achieved due to the uncertaintly principle 
- the collisions induce uncertainty of momenta of the scattered sub-particles~\citep{Sbitnev2013a}.

 Next we proclaim that the pressure $p$ consists of sum of two pressures, $p_{1}$ and $p_{2}$. And the pressure $p_{1}$ follows from the Fick's laws. 
The first Fick's law says that the diffusion flux, ${\mathit j}_{\mu}$, is proportional to the negative value of the density mass gradient~(\ref{eq=4}).
One can see
 that the four-laplacian $D {\widetilde\partial}^{\,\mu}{\mathit j}_{\mu}
= -D^2\, {\widetilde\partial}^{\,\mu} {\widetilde\partial}_{\mu}\rho_{_{M}}$ 
has the dimension of the pressure. So, we write
\begin{equation}
p_{1} = -D^{2}
\biggl(
  \nabla^2 \rho_{_{M}}    -    {{\seltat}\over{c^2}}{{\partial^2\;}\over{\partial t^2}}\rho_{_{M}} 
\!\biggr).
\label{eq=7}
\end{equation}
As for the pressure  $p_{2}$, one can note first that the kinetic energy of the diffusion flux is 
$(m/2)({\mathit j}^{\,\mu}{\mathit j}_{\mu}/\rho_{_{M}}^{2})$.
It means that there exists one more pressure as the kinetic energy  per unit volume of the medium in question:
\begin{equation}
 p_{2} =  {{D^2}\over{2\rho_{_{M}}}}
  \biggl(\!
     (\nabla \rho_{_{M}})^{2}
 -        \biggl( 
               {{\deltat}\over{c}}{{\partial}\over{\partial t}}\rho_{_{M}}
          \biggr)^{\!2\,}
   \biggr).
\label{eq=8}
\end{equation}

 Now we can see that sum of the two pressures, $p_1 + p_2$, divided by $\rho$ (we remark that $\rho_{_{M}} = m\rho$)  reduces to the quantum potential
\begin{eqnarray}
\nonumber
   Q &=& {{p_{1}+p_{2}}\over{\rho}} =
-{{\,\hbar^{2}}\over{4m\rho}}
\biggl(
        \nabla^2 \rho    -   { {{\seltat}\over{c^2}} {{\partial^2\;}\over{\partial t^2}}}\rho  
\biggl)
\\ \nonumber
&&\hspace{54pt}
+{{\,\hbar^{2}}\over{8m \rho^2}}
\biggr(\!
   \bigr(
          \nabla \rho
    \bigl)^{\!2}
-   \biggl( 
           {{\deltat}\over{c}} {{\partial}\over{\partial t}}\rho
    \biggr)^{\!2}
\biggr)
\\
=&-&{{\,\hbar^2}\over{2mR}}
\biggl(
   \nabla^2 - {{\seltat}\over{c^2}} {{\partial^2\;}\over{\partial t^2}}
\biggr) R 
 = - {{\hbar^2}\over{2m}}{{{\partial}_{\mu}{\partial}^{\,\mu}R}\over{R}}.
 \hspace{24pt}
\label{eq=9}
\end{eqnarray}
 Here $R = \sqrt{\rho}$.
  When ${\mathit v}=1$ the quantum potential 
 is in complete concordance with that given in~\citep{Nikolic2007}.
On the other hand, when ${\mathit v}\rightarrow0$
this potential degenerates to the non-relativistic one.

\subsection{\label{subsect2B}Scalar field in a general case}

 Return to Eq.~(\ref{eq=3}).  
Since the tensor $T^{\mu\nu}$ is symmetric, one can believe that it comes from the scalar field.
After multiplying this equation by $\eta^{\mu\nu}\eta_{\nu\mu}=1$
 and taking into account that
${\partial}_{\mu}\eta^{\mu\nu}\eta_{\nu\mu}{\mathit u}^{\mu}{\mathit u}^{\nu}
={\partial}^{\,\nu}{\mathit u}_{\,\nu}{\mathit u}^{\nu}$
 we have
\begin{equation}
\hspace{-8pt}
  \biggl(\!{{\epsilon + p}\over{\rho}}\gamma\!\biggr) {\partial}^{\,\nu}{\mathit u}_{\,\nu}{\mathit u}^{\,\nu}
 \!-\! {\partial}^{\,\nu} \biggl(\!{{\epsilon + p}\over{\rho}}\gamma\!\biggr)
 + {\partial}^{\,\nu} Q = 0.
\label{eq=10}
\end{equation}
In front of the second member we write minus
 inasmuch as ${\mathit u}_{\,\nu}{\mathit u}^{\nu}=-1$.

Keeping in mind the Klein-Gordon equation in perspective 
we first multiply Eq.~(\ref{eq=10}) by 2 and suppose that
\begin{equation}
  mc^2 = 2{{(\epsilon + p)}\over{\rho}}\gamma.
\label{eq=11}
\end{equation}
The mass $m$ hereinafter is $m_{0}\gamma$, where $m_{0}$ is the rest mass and $\gamma = (1-v^2)^{-1/2}$.
Also  we write the letter $c$ meaning the speed of light, which is adopted as 1 according to our convention.
Inasmuch as the superfluid medium is irrotational, the first term of Eq.~(\ref{eq=10})
 is rewritten as ${\partial}^{\,\nu}{\mathit u}_{\,\nu}{\mathit u}^{\,\nu}
= {\mathit u}^{\,\nu}{\partial}^{\,\nu}{\mathit u}_{\,\nu}
+ \eta^{\nu\mu}{\mathit u}_{\,\mu}{\partial}^{\,\nu}\eta_{\nu\mu}{\mathit u}^{\,\mu}
={\partial}^{\,\nu}{\mathit u}^{2}$. 
The solution of Eq.~(\ref{eq=10}) is as follows
\begin{equation}
   m{\mathit v}^{2} - mc^2 +2Q = 2C.
\label{eq=12}
\end{equation}
 Here ${\mathit v}_{\mu}={\mathit u}_{\mu}c$ and $C$ is the integration constant
having dimension of energy.
Since the superfluid medium under consideration is irrotational, we can express the velocity ${\mathit v}_{\mu}$ through the gradient of the scalar field $S+\hbar\omega\gamma^{-1}t$. 
The last extra term vanishes as $\gamma^{-1}$ tends to zero (the relativistic sector),
but this term will be important in the non-relativistic limit when getting the Schr\"odinger equation.
So, the velocity ${\mathit v}_{\mu}$ can be expressed as follows
\begin{eqnarray}
\nonumber
&&  {\mathit v}_{\mu} = -{{1}\over{m}} {\partial}_{\mu}(S+E{\gamma}^{-1}t)
= -{{1}\over{m}} {\partial}_{\mu}S - {{E}\over{mc\gamma}}, \\
\nonumber
&&
  {\mathit v}^{2} = {{1}\over{m^2}}{\partial}_{\mu}\,(S+E{\gamma}^{-1}t){\partial}^{\,\mu}(S+E{\gamma}^{-1}t) \\
&=&
  {{1}\over{m^2}}{\partial}_{\mu}\,S{\partial}^{\,\mu}S
+  {{2E}\over{\gamma m^2c}}{\partial}_{\,0}\,S
+  {{E^2}\over{\gamma^{2}m^2c^2}}.
\label{eq=13}
\end{eqnarray}
 Here we need to say some words about the phase $\omega\gamma^{-1}t$.
 It comes from the expression ${\Delta t} = {\Delta t_{_{0}}}\gamma$ describing the time dilation.
This phase belonging to the unitary group $U(1)$ carries out the timekeeping of  the path $tc$~\citep{Poluyan2005}.
 It is seen that the timekeeping tends to zero as $\gamma^{-1}$ tends to zero, that is, as  ${\mathit v}\rightarrow 1$.
The energy of the timekeeping is defined as $E = \hbar\omega = m_{0}c^2$. Where $m_{0}$ is the rest mass. Its contribution disappears  as the velocity approaches the speed of light. 

Now multiplying  Eq.~(\ref{eq=12}) by $m$  we  rewrite it in details
\begin{eqnarray}
\nonumber
&&
\underbrace{ \partial_{\mu} S \partial^{\,\mu} S}_{\rm (a)}  +
\underbrace{ {{2E}\over{c\gamma}}\partial_{\,0} S + {{E^2}\over{c^2{\gamma}^2}} }_{\rm (b)}
\\
&&
   -   m^2c^2   -  {\hbar^2}\, {{\partial_{\mu} \partial^{\,\mu} R }\over{R}} = 2mC.
\label{eq=14}
\end{eqnarray}
Let us add to this equation for pair also the continuity equation~\citep{Fiscaletti2012} 
\begin{equation}
  \partial_{\,\mu} ( \rho\,\partial^{\,\mu}S ) = 0.
\label{eq=15}
\end{equation}
 
The parameter $\gamma = (1-{\mathit v}^{2})^{-1/2}$ presented in the denominators of the two terms 
 enclosed in the brace (b) in Eq.~(\ref{eq=14}) determines transition to either the relativistic limit or 
to the non-relativistic limit.

\subsubsection{\label{subsubsect2B1}Relativistic sector}

 At relativistic limit, ${\mathit v}\rightarrow 1$, $\gamma$ tends to infinity.
 In this case the terms enclosed in the brace (b) in Eq,~(\ref{eq=14}) vanish and we have
\begin{equation}
 \partial_{\mu} S \partial^{\,\mu} S    -   m^2c^2   -  {\hbar^2}\, {{\partial_{\mu} \partial^{\,\mu} R }\over{R}}
 = 2mC.
\label{eq=16}
\end{equation}
 The both equations, Eqs.~(\ref{eq=15}) and~(\ref{eq=16}), 
are extracted from the Klein-Gordon equation
\begin{equation}
 \partial_{\mu}\partial^{\,\mu} \psi + 
 {{m^2c^2}\over{\hbar^2}}\psi + {{2m}\over{\hbar^2}}C\psi = 0
\label{eq=17}
\end{equation}
 as soon as we substitute in this equation the wave function $\psi$ represented in the polar form
\begin{equation}
  \psi = R \exp\{ {\bf i}S/\hbar \}
\label{eq=18}
\end{equation}
 and separate the solutions into the real, Eq.~(\ref{eq=16}), and imaginary, Eq.~(\ref{eq=15}), parts.
 Here $R=\rho^{1/2}$ is the amplitude of the wave function and $S/\hbar$ is its phase.
The probability density $\rho({\vec r},t) = R(\vec r,t)^2$ shows distribution of the virtual particles in the vicinity of the 4-point $(t,{\vec r})$. Whereas the action $S({\vec r},t)$ characterizes a degree of mobility of these particles in the same place.

 Computations done above give a good concurrence with those
 that have been fulfilled by Fiscaletti~\citep{Fiscaletti2012}. In his work the quantum Hamilton-Jacobi
equation has a view
\begin{equation}
   \partial_{\mu} S \partial^{\,\mu} S = m^2c^2 (1 + {\widetilde Q})
\label{eq=19}
\end{equation}
with the quantum potential defined as
\begin{equation}
 {\widetilde Q} = {{\hbar^2}\over{m^2c^2}}{{1}\over{R}}
 \biggl(
   \nabla^{2} - {{1}\over{c^2}}{{\partial^2\,}\over{\partial\,t^2}}
 \biggr) R.
\label{eq=20}
\end{equation}\\
Comparing Eqs.~(\ref{eq=16}) and~(\ref{eq=19})  we find their good concurrence when $C = 0$. 
The difference in the quantum potentials~(\ref{eq=9}) and~(\ref{eq=20})
 is due to their  individual  definitions.
The potential $Q$ has the dimension of energy, whereas ${\widetilde Q}$ is dimensionless due to dividing by $mc^2$.
Opposite signs at the quantum potentials
 are due the choice of the metric tensors, $\eta_{\mu\nu}$ or $g_{\mu\nu}$, having the opposite signatures 
 $(- + + +)$ and $(+ - - -)$, respectively.

\subsubsection{\label{subsubsect2B2}Non-relativistic sector}

At non-relativistic limit ${\mathit v}\rightarrow 0$. It means that $\gamma$ converges to unit.
 Observe that in this limit the third and fourth terms in Eq.~(\ref{eq=14}) kill each other, $E^2/c^2 - m_{0}^2c^2 = 0$.
First we rewrite the term enclosed in the brace (a) in Eq.~(\ref{eq=14}) in the form
${\partial_{\mu}}S{\partial^{\mu}}S = {\partial_{\,0}}S{\partial^{\,0}}S +(\nabla S)^2$, where 
$\partial_{\,0} = \partial/\partial\tau$ and $\tau = ct$. 
So, we need to compare $\partial S/\partial t$ and $E = m_{0}c^2$.
In the non-relativistic limit we have the following inequality $m_{0}c^2 \gg \partial S/\partial t$.  
So we can rewrite Eq.~(\ref{eq=14}) as follows
\begin{equation}
{{\partial }\over{\partial\,t}}S + {{1}\over{2m_{0}}}(\nabla S)^2
-{{\hbar^2}\over{2m_{0}}}{{\nabla^{\,2}R}\over{R}} = C.
\label{eq=21}
\end{equation}

 This equation is the modified Hamilton-Jacobi equation because of the quantum potential loaded to it.
 This equation is supplemented by the continuity equation~\citep{Fiscaletti2012}:
\begin{equation}
 {{\partial\,\rho}\over{\partial\,t}}
 + \nabla \biggl(
                          \rho\,{{\nabla S}\over{m_{0}}}
                \biggr) = 0.
\label{eq=22}
\end{equation}
  Here the density distribution $\rho = R^2$
 and $\nabla S/m_{0} = {\vec{\mathit v}}$ is the current velocity.
 These equations, Eqs.~(\ref{eq=21}) and~(\ref{eq=22}), are extracted from
 the following Schr\"odinger equation
\begin{equation}
  {\bf i}\hbar {{\partial\,\psi}\over{\partial\,t}} =
- {{\hbar^{2}}\over{2m_{0}}}\nabla^{2} \psi  -C \psi.
\label{eq=23}
\end{equation}
 By substituting into this equation the wave function $\psi$ represented in the polar form~(\ref{eq=18})
and separating on real and imaginary parts we come to Eqs.~(\ref{eq=21}) and~(\ref{eq=22}), respectively.

\section{\label{sect3}Relativistic non-perfect fluid:
   vortex tubes and others}

In previous chapter we considered the perfect relativistic fluid. It represents, in fact, the superfluid medium. By adopting it as the physical vacuum we have derived the quantum equations in assumption of existence of the scalar field $S$ gradient of which gives the fluid velocity. Next let us add to 
the energy-momentum tensor~(\ref{eq=1})
 a term describing the viscosity of the fluid~\citep{RezzollaZanotti2013}
\begin{equation}
\hspace{-16pt}
 T^{\,\mu\nu} =  (\epsilon + p){\mathit u}^{\mu}{\mathit u}^{\nu} + p \eta^{\,\mu\nu}  + {\Pi}^{\mu\nu} 
\label{eq=24}
\end{equation}
 A new  term ${\Pi}^{\mu\nu}$ embedded in this equation is the viscous stress tensor~\citep{Romatschke2010a}.
In the general case we represent it in the following view~\citep{RezzollaZanotti2013}
\begin{equation}
 \Pi^{\mu\nu} =
\mu c(\partial^{\mu}{\mathit u}^{\nu} + \partial^{\nu}{\mathit u}^{\mu})
  + c\biggl(
             \zeta - {{2}\over{3}}\mu
     \biggr)
     \partial^{\,\mu}{\mathit u}_{\mu}\eta^{\,\mu\nu}.
\label{eq=25}
\end{equation}
Since the dimensionality of the viscosity coefficients, shear $\mu$ and bulk $\zeta$, is Pa$\cdot$s and the dimension of $c\,\partial^{\mu}{\mathit u}^{\nu}$ is s$^{-1}$, the dimension of $\Pi^{\mu\nu}$ is Pa = Joule/m$^3$.

We shall consider the incompressible fluid. It means that $\nabla\rho=0$.
The conservation law of the energy-momentum tensor takes the form
\begin{eqnarray}
\nonumber
\hspace{-18pt}
   \partial_{\mu} \biggl({{T^{\mu\nu}}\over{\!\rho}}\biggr) 
& =&
 \partial_{\mu}  \biggr(\!{{\epsilon+p}\over{\!\rho}}\gamma {\mathit u}^{\mu}{\mathit u}^{\nu} \!\biggl)
 \\
\hspace{-18pt}
&+& \partial^{\,\nu} Q
\;+\; \partial_{\mu} (\mu(t)/\rho)  \pi^{\mu\nu}     = 0.
\label{eq=26}
\end{eqnarray}
 Here $Q=p/\rho$ is the quantum potential, due to which transition to the quantum equations can be assured.
The term $\pi^{\mu\nu}$ has a view
\begin{equation}
  \pi^{\mu\nu} =
c (\partial^{\mu}{\mathit u}^{\nu} + \partial^{\nu}{\mathit u}^{\mu})
              - c{{2}\over{3}}
     \partial^{\,\mu}{\mathit u}_{\mu}\eta^{\,\mu\nu}.
\label{eq=27}
\end{equation}
 For the sake of simplicity, we take further into account only the shear viscosity.
Note that the viscosity we believe is a function of time~\citep{Sbitnev2015b, Sbitnev2015c}.
 We suppose that the viscosity average on time vanishes.
 But its variance is non-zero.
 It means that the parameter $\mu(t)$ is a fluctuating quantity about zero.
 This supposition is a very impressive.
We hope to obtain long-lived vortices having non-zero vorticity's  cores~\citep{Sbitnev2015b, Sbitnev2015c}.
But first let us show that the equation $\partial_{\mu}(T^{\mu\nu}/\rho)=0$ can be reduced to the relativistic 
Navier-Stokes equation. Here we repeat the computations of J.~W.~van~Holten~\citep{Holten2006}.

\subsection{\label{subsect3A}Relativistic form of the Navier-Stokes equation}

Let us split Eq.~(\ref{eq=26})  into the time and space components.
 The equation with $\nu=0$ becomes:
\begin{eqnarray}
\nonumber
&&  {{\partial}\over{\partial\tau}}\biggl(\!{{\epsilon+p}\over{\rho}}\gamma\!\biggr)
  + \nabla\!\cdot\!\biggl(\!
                            {{ \epsilon + p } \over {\rho }}\gamma  {\vec{\mathit v}}
                           \!\biggr) \\
 &+&  {{\partial\,Q}\over{\partial\tau}}
 +{{\partial (\mu(t)/\rho)}\over{\partial\tau}}  \cdot \pi^{0,0}
 + {{\mu(t)}\over{\rho}} \partial_{\mu}\pi^{\mu, 0}
  = 0,
\hspace{22pt}
\label{eq=28}
\end{eqnarray}
 whilst the space components with $\nu=i$ becomes
\begin{eqnarray}
\nonumber
 && 
{{\partial\;}\over{\partial \tau}}
            \biggl(
                    {{\epsilon+p}\over{\rho}}\gamma {\mathit v}_{i}
             \biggr) 
 + \nabla_{}\!\cdot\!
             \biggl(
                     {{\epsilon+p}\over{\rho }}\gamma {\vec{\mathit v}}\!\cdot\!{\mathit v}_{i}
             \biggr) 
\\ 
&+&\nabla_{i} Q +
 {{\partial(\mu(t)/\rho)}\over{\partial\tau}}  \cdot \pi^{0,i}
 + {{\mu(t)}\over{\rho}} \partial_{\mu}\pi^{\mu,i}
 = 0.
\hspace{22pt}
\label{eq=29}
\end{eqnarray}
 Here $\tau = ct$ and $i$ runs $1, 2, 3$.
 By replacing the derivatives of $(\epsilon+p)\gamma/\rho$ and $(\epsilon+p)\gamma {\vec{\mathit v}}/\rho$
 with the time and the direction  with relevant terms taken from Eq.~(\ref{eq=28})
 we obtain the following expression:
\begin{eqnarray}
\nonumber
&& 
 {{\epsilon+p}\over{\rho}}\gamma
\biggl(\!
  {{\partial {\vec{\mathit v}}}\over{\partial \,\tau}} + ({\vec{\mathit v}}\!\cdot\!\nabla{\vec{\mathit v}})
\biggr) 
+  \nabla_{}Q 
 -  {\vec{\mathit v}} {{\partial\;}\over{\partial \,\tau}} Q
\\ \nonumber
 &&+ {{\partial(\mu(t)/\rho)}\over{\partial\tau}} 
 \cdot (\pi^{0,i}  -  {\vec{\mathit v}}\, \pi^{0,0}) \\
&&
 + {{\mu(t)}\over{\rho}} (\partial_{\mu}\pi^{\mu,i}  -  {\vec{\mathit v}}\, \partial_{\mu}\pi^{\mu, 0})  = 0.
\label{eq=30}
\end{eqnarray}
 Recall that $\gamma = (1-{\mathit v}^2)^{-1/2}$.
 When ${\mathit v}$ tends to 1 the parameter $\gamma$ goes to infinity.
And this equation describes the fluid motion with relativistic velocities.
 Here we define
\begin{equation}
  m c^2 =  {{\epsilon+p}\over{\rho}}\gamma,
\label{eq=31}
\end{equation}
 where the relativistic mass $m$ is $m_{0}\gamma$, and $m_{0}$ is the rest mass.

 When ${\mathit v}$ tends to 0 the parameter $\gamma$ goes to 1, however.
It is the non-relativistic limit. Let us look into this limit.

\subsubsection{\label{subsect3A1}Non-relativistic limit}

 At non-relativistic limit, ${\mathit v}\rightarrow 0$, $\gamma$ converges to unit.
In this case $mc^{2}$ reduces to $m_{0}c^2$, where  $m_{0}$ is the rest mass.
Let the multiplier $c^2$ be shifted inside brackets in the first line of Eq.~(\ref{eq=30}) 
\begin{equation}
 m_{0} c^2
\biggl(\!
  {{\partial {\vec{\mathit v}}}\over{\partial \,\tau}} + ({\vec{\mathit v}}\!\cdot\!\nabla{\vec{\mathit v}})
\!\biggr)
=
 m_{0}
\biggl(\!
  {{\partial c{\vec{\mathit v}}}\over{\partial \,t}} + (c{\vec{\mathit v}}\cdot\!\nabla{c\vec{\mathit v}})
\!\biggr) 
\label{eq=32}
\end{equation}
and call the term $c{\vec{\mathit v}}$ as the real velocity, ${\vec{\mathit V}}$, of the non-relativistic fluid.
Taking into account that the third and four terms in Eq.~(\ref{eq=30}) vanish in the non-relativistic limit
we come to the non-relativistic form of the Navier-Stokes equation
\begin{equation}
  m_{0}
\biggl(\!
  {{\partial {\vec{\mathit V}}}\over{\partial \,t}} + ({\vec{\mathit V}}\cdot\!\nabla{\vec{\mathit V}})
\!\biggr) 
= \nabla Q
+ {{\mu(t)}\over{\rho}}\nabla^{2}  {\vec{\mathit V}}.
\label{eq=33}
\end{equation}
 For the sake of simplicity we cut the viscosity term up to 
$(\mu(t)/\rho)\partial_{\,i}\partial^{\,i} c{\mathit v}^{i} \Rightarrow (\mu(t)/\rho)\nabla {\vec{\mathit V}}$,
as it is done for ease of analytical studies~\citep{Sbitnev2015b, Sbitnev2015c}.
 
Observe that the term $ ({\vec{\mathit V}}\cdot\!\nabla{\vec{\mathit V}})$
can be rewrite as follows
\begin{equation}
 ({\vec{\mathit V}}\cdot\!\nabla{\vec{\mathit V}}) 
 = \nabla {\mathit V}^2/2 
 + [{\vec\omega}\times{\vec{\mathit V}}].
\label{eq=34}
\end{equation}
 The first term multiplied by the  mass $m_{0}$ is gradient of the  kinetic energy.
 Whilst the second term multiplied by the same mass describes the vortex force.
 Here ${\vec\omega}=[\nabla\times{\vec{\mathit V}}]$ is named the vorticity.
Obviously, from here it follows that the velocity ${\vec{\mathit V}}$ is represented by superposition
of irrotational and solenoidal components~\citep{KunduCohen2002}.
 The first component stems from the gradient of the scalar field.
While the second component is due to the vector field where the Kelvin-Stokes theorem gives
the rule for calculus of the curl of the vector field over some surface~\citep{Katz1979}.

By applying the curl operator to Eq.~(\ref{eq=33}) we get the equation for the vorticity
\begin{equation}
  {{\partial{\vec\omega}}\over{\partial\,t}}
 + ({\vec\omega}\cdot\nabla){\vec{\mathit V}}
 = \nu(t) \nabla^{\,2}{\vec\omega}.
\label{eq=35}
\end{equation}
 Here $\nu(t) = \mu(t)/(m_{0}\rho)$ is the kinetic viscosity. Its dimension is [length$^2$/time].

 Assuming that the fluid is a physical vacuum, which is superfluid~\citep{Volovik2003}, one says that the viscosity is missing.
 In such a case, the vorticity $\vec\omega$ is concentrated in the point of the origin. Its mathematical representation is a $\delta$-function. This singularity can be a source of possible divergences of computations in later on.

Here we proclaim that the viscosity coefficient, $\nu(t)$,  
 accidentally changes sign with the lapse of time.
Let, at that, the average in time  of the coefficient $\nu(t)$ be equal to zero, but its variance stays positive. 
Such unusual changes of sign in the viscosity  can mean only that there occurs
 the exchange of the energy of the vortex
 with zero fluctuations of the vacuum~\citep{Sbitnev2015b, Sbitnev2015c}.

 Further we shall consider a simple model of the vortex. Let us look into the vortex tube in its cross-section which is oriented along the $z$-axis  and its center is placed in the coordinate origin of the plane $(x, y)$. Eq.~(\ref{eq=35}), written down in the cross-section of the vortex, is as follows
\begin{equation}
  {{\partial\, {\omega}}\over{\partial\,t}} =
 \nu(\,t\,)\Biggl(
    {{\partial^{\,2}\omega}\over{\partial\,r^{2}}}
 +{{1}\over{r}}{{\partial\,\omega}\over{\partial\,r}}
               \Biggr).
\label{eq=36}
\end{equation}
 The solution of this equation reads
\begin{equation}
   \omega(r,t) =
{{\mit\Gamma}\over{  \Sigma(\,t\,)  }}
\cdot \exp\Biggl\{
  - {{r^2}\over{  \Sigma(\,t\,)  }}
  \Biggr\}
\label{eq=37}
\end{equation}
 and the orbital speed is
\begin{eqnarray}
\nonumber
 {\mathit{ V}}(r,\,t) 
&=& {{1}\over{r}}\int\limits_{0}^{r}\omega(r',t)r'dr' \\
&=&  {{\mit\Gamma}\over{2 r}}\Biggl(\!1 - 
 \exp{\Biggl\{
   - {{r^2}\over{  \Sigma(\,t\,)  }}
  \Biggr\}}\!
  \Biggr).
\label{eq=38}
\end{eqnarray}
 Here $\mit\Gamma$ is the integration constant having dimension m$^2$/s
 and the denominator $\Sigma(\,t\,)$ reads
\begin{equation}
  \Sigma(\,t\,) = 4 \Biggl( \int\limits_{0}^{t} \nu(\,t'\,)dt' + \sigma^{2} \Biggr) 
\label{eq=39}
\end{equation}
 Here $\sigma$ is an arbitrary constant such that the denominator is always positive.
\begin{figure}[htb!]
  \centering
  \begin{picture}(200,125)(22,20)
      \includegraphics[scale=0.75]{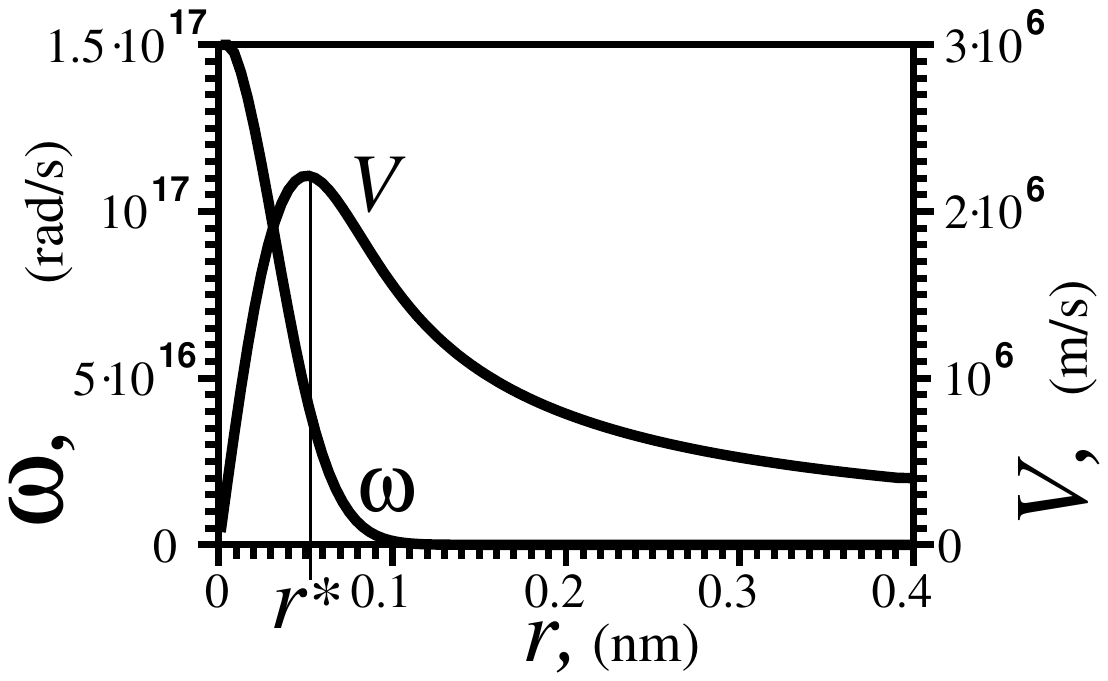}
  \end{picture}
  \caption{
  Vorticity $\omega(r)$ and orbital speed ${\mathit V}(r)$  as functions of $r$ at given $t$
  for ${\mit\Gamma}=\pi\cdot10^{-4}$ m$^2$/s, $\nu=\hbar/2m_{0}$ ($m_{0}$ is the electron mass),
 $\Omega=6.58\cdot10^{15}$ s$^{-1}$, and $\sigma=1.914\cdot10^{-10}$ m. 
 The  boundary of the core is $r^* \approx 5.29\cdot10^{-11}$ m.
   }
  \label{fig=1}
\end{figure}

 For the sake of simplicity we set
\begin{equation}
  \nu(\,t\,) = \nu\cos(\Omega\,t )
=\nu{{e^{{\bf i}\Omega\,t}+e^{-{\bf i}\Omega\,t}}\over{2}}
\label{eq=40}
\end{equation}
In this case
\begin{equation}
  \Sigma(\,t\,) = 4 ({{\nu}/{\Omega}})\sin(\Omega\,t) + \sigma^{2} .
\label{eq=41}
\end{equation}
Here $\Omega$ is an oscillation frequency and $\nu$ let be equal to $\hbar/2m_{0}$, compare with Eq.~(\ref{eq=6}).
 Behavior of the functions~(\ref{eq=37}) and~(\ref{eq=38})
 calculated for the case of the electron-positron pairs fluctuating on the first Bohr orbit
 are shown in Fig.~\ref{fig=1}.
 
The functions $\omega(r,t)$ and ${\mathit V}(r,t)$ oscillate about  some average values
 limited by $\sigma$~\citep{Sbitnev2015b, Sbitnev2015c}.
 It looks as the vortex is  pulsating in time, at least pulsating the core's wall 
 marked in the figure by the radius $r^{*}$.
 The pulsations are due to the annihilation-creation process of the virtual electron-positron pairs.
\begin{figure}[htb!]
  \centering
  \begin{picture}(200,170)(25,10)
      \includegraphics[scale=0.45]{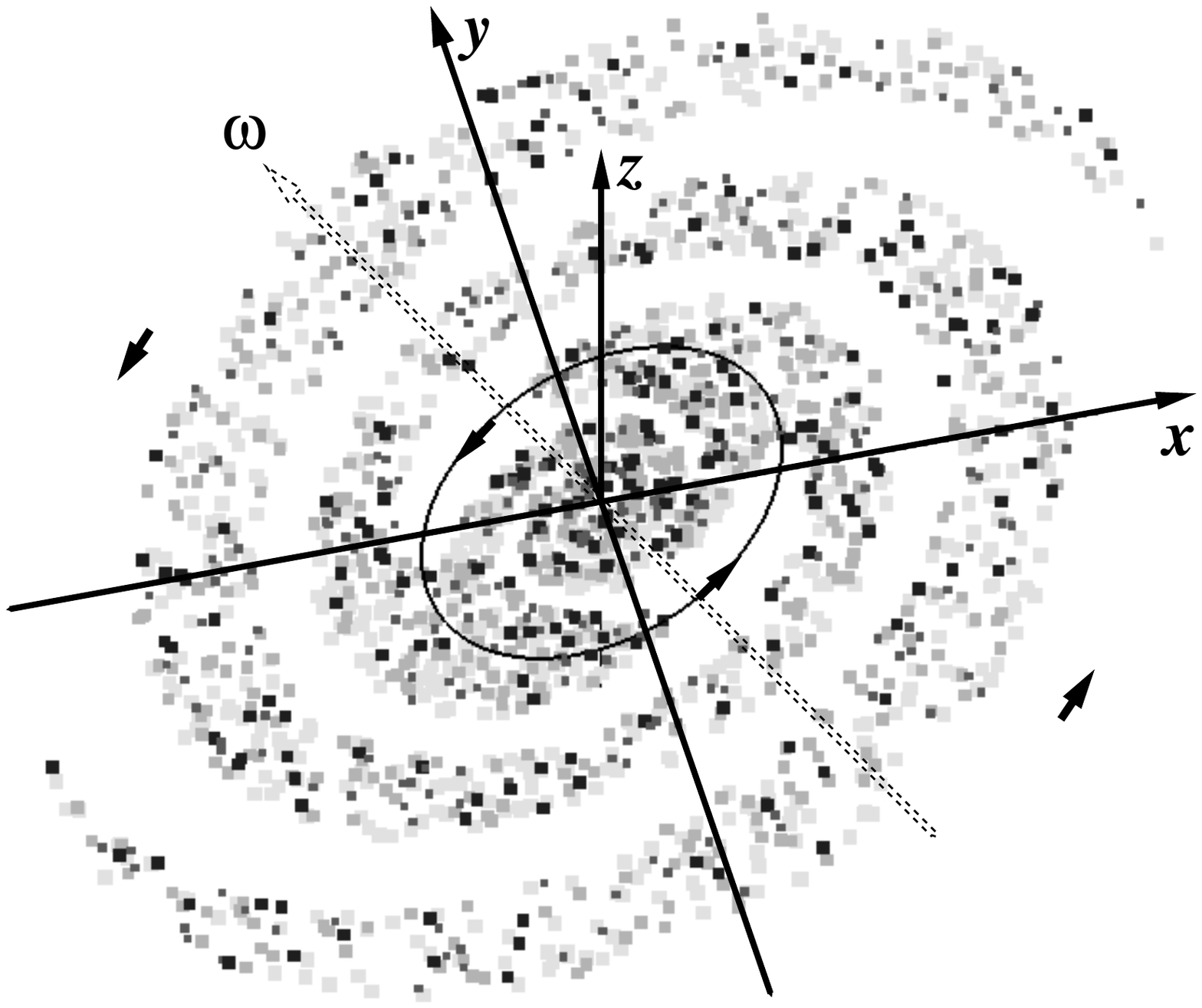}
  \end{picture}
  \caption{
   Vortex's simulation. 
  Circle drawn in the center covers an area of the core.
  On edge of the core the orbital speed reaches a maximal value.
  Rotation is shown by arrows.
 The dotted arrow being perpendicular to the rotation plane
 points orientation of the vorticity.
   }
  \label{fig=2}
\end{figure}
\begin{figure}[htb!]
  \centering
  \begin{picture}(200,125)(20,15)
      \includegraphics[scale=0.75]{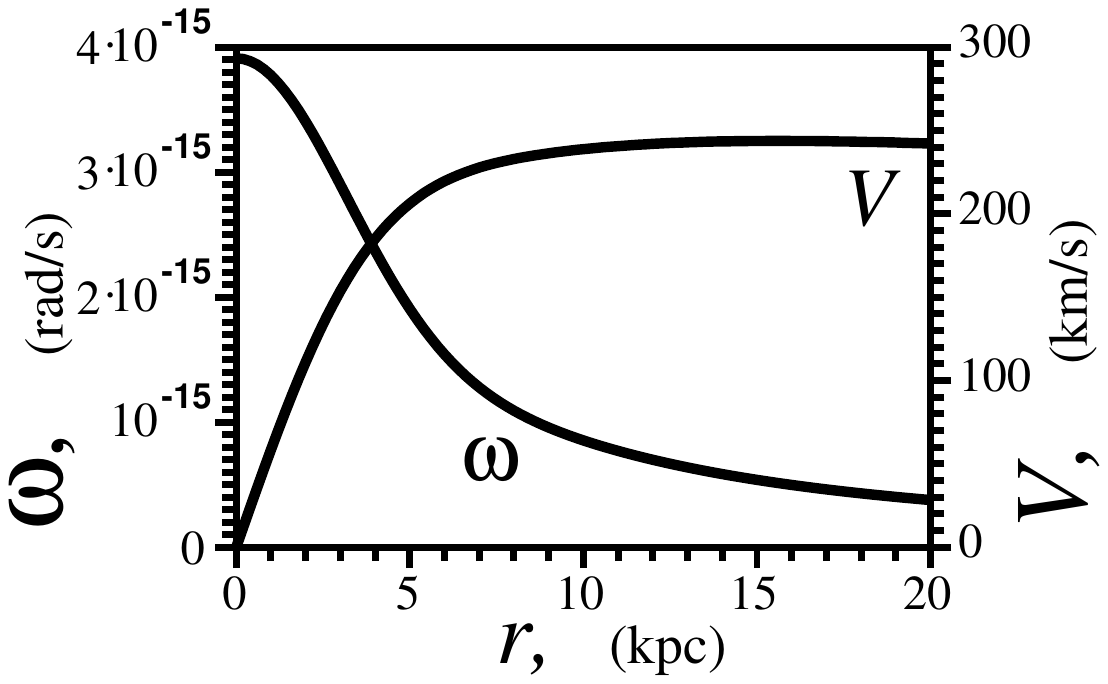}
  \end{picture}
  \caption{
  Vorticity $\omega(r)$ and orbital speed ${\mathit V}(r)$  as functions of $r$ at given $t$.
 Scaling  units are adopted as stated in~\citep{deBlokEtAl2001}. 
   }
  \label{fig=3}
\end{figure}

\section{\label{sect4}Vortex strings and dark matter}

Simulation of the vortex is shown in Fig.~\ref{fig=2}.
Simulating the vortex we choose a two-arm mode in the azimuthal equidistant projection
 because of which the vortex manifests a good spiral form. It looks almost like a galaxy.
 The vortex has a core which does not exceed the radius~$r^{*}$
 where the point of inflection of the orbital velocity ${\mathit V}(r)$ occurs.
 In the figure this area is outlined by a circle with the arrows pointing out the rotation of the matter.
 Below this circle the velocity diminishes and in the core we see agglomeration of the slowly rotating matter.
Above this circle the velocity reaches, at least, its maximal values and stays almost constant.
 
It is that the stars in the galaxy rotate around the galactic core with almost constant speed even being located far from the galactic core.
Really, the orbital speed of stars and gas, ${\mathit V}$, is rising inside the core and reach almost constant level outside the core~\citep{deBlokEtAl2001}.
The orbital speed can look as shown in Fig.~\ref{fig=3}.

We assume that there is a wide spectrum of the viscosity coefficients. 
For the sake of simple presentation we believe that the spectrum is discrete with equidistant position
of each component which has a form
\begin{equation}
  \nu_{\,n}(\,t\,) = {{c^{\,2}}\over{\Omega_{n}}}\cos(\Omega_{n}t).
\label{eq=42}
\end{equation}
Take yourself attention that the coefficient $c^{\,2}/\Omega_{n}$ is fully equivalent to 
the diffusion coefficient~(\ref{eq=5}), here $c$ is the light speed.
Also one can see that $c^{\,2}/\Omega_{n}$ submits to the $1/f$-law 
that exhibits itself as the flicker-noise (colored noise)~\citep{HanggiJung1995} when $n$ tends to infinity.
Namely, we admit that
\begin{equation}
  \Omega_{n} \sim n^{-\alpha}, ~~~~ \alpha \ge 1.
\label{eq=43}
\end{equation}
It should take into attention here, that the spectrum  is condensed in the point $\Omega = 0$.
In other words, the strongest contribution to the vorticity give modes with frequencies close to zero:
\begin{equation}
 \Sigma_{n}(\,t\,) = 4 \biggl(
  {{c^{\,2}}\over{\Omega_{n}^{2}}}\sin(\Omega_{n}t) + \sigma_{n}^{2}
                                \biggr).
\label{eq=44}
\end{equation}

 The vorticity
\begin{equation}
 \omega(r,t) = {{\mit\Gamma}\over{N}} \sum\limits_{n=1}^{N}
 {{1}\over{ \Sigma_{n}(\,t\,)}}\exp\Biggr\{
                                                                                    -{{r^2} \over {\Sigma_{n}(\,t\,)}}
                                                                         \Biggl\}
\label{eq=45}
\end{equation}
 and the orbital speed
\begin{equation}
 {\mathit V}(r,t) = {{\mit\Gamma}\over{2 rN}}
 \sum\limits_{n=1}^{N}
 \Biggr(
 1 -\exp\Biggr\{
                      -{{r^2} \over {\Sigma_{n}(\,t\,)}}
             \Biggl\}
 \Biggl).
\label{eq=46}
\end{equation}
 These curves are calculated   for  choosing 
 ${\mit\Gamma} = 10^{27}$~m$^2$/s,
 $\Omega_{n} = 10^{-11}n^{-1}$~s$^{-1}$, and $N=25$. The parameter $\sigma_{n} = 4c/\Omega_{n}$ m
  ranges from 10,000 to 300,000 light years which covers  the diameter of the ordinary spiral galaxy~\citep{Howell2015}.
 This parameter  tends to infinity as $\Omega$ goes to zero.
 It prevents a catastrophe of the barion matter with the course of time.
 The curves are shown in Fig.~\ref{fig=3}. 

\begin{figure}[htb!]
  \centering
  \begin{picture}(200,230)(20,10)
      \includegraphics[scale=0.6]{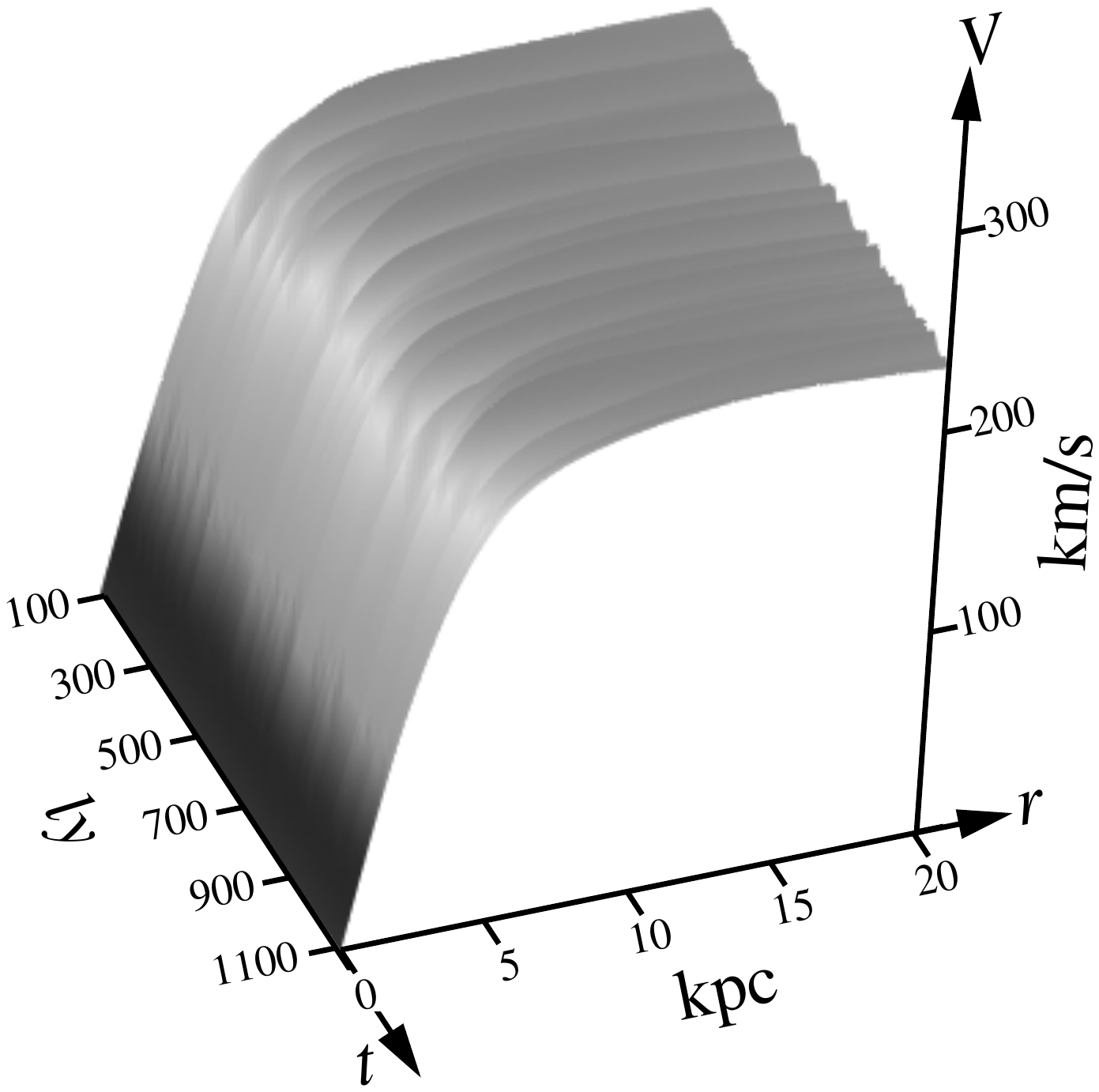}
  \end{picture}
  \caption{
  Orbital speed ${\mathit V}$  as function of 
 the radius $r$ from the galactic center (in kiloparsec) 
 and time $t$ (in light years).
   }
  \label{fig=4}
\end{figure}

We have shown that the orbital speed can manifest a roughly flat rotation curve if the vorticity bundles are assembled according to the $1/f$-law. The frequency range in that case stretches from high frequencies belonging to the quantum realm up to about $6\cdot10^{-14}$ Hz.
These ultra-low frequencies, in terms of wavelengths, are covering almost entirely the diameter of the ordinary spiral galaxy, $\lambda=c/{\omega} \approx 5\cdot10^{\,5}$ light years, or about 150 kpc.
In this sense, the formulas~(\ref{eq=45}) and~(\ref{eq=46}) describe thermalization of the vorticity and angular velocities of spiral galaxies as a result of the energy exchange of baryonic matter with the physical vacuum.

Fig.~\ref{eq=4} shows evolution of the orbital speed,~${\mathit V}$, for the time ranging from 100 to 1100 light years. Through the whole time interval the orbital speed keeps its own form. One can see, however, that the curve experiences small fluctuations in  time, resembling the breathing of the galaxy. This breathing is caused by the exchanging energy with the physical vacuum on the ultra-low frequencies owing to the term $\Sigma_{n}(\,t\,)$. It manifests in variations of the vorticity $\omega$.
 On these ultra-low frequencies this exchange occurs by the virtual particle-antiparticle pairs named as  gravitational dipoles~\citep{Hajdukovic2011a, Hajdukovic2011b, Hajdukovic2014a}. 
It is believed that such a manifestation stems from dark matter~\citep{Rubin2004}.
But  Hajdukovic writes~\citep{Hajdukovic2011a}: "dark matter does not exist but is an illusion created by the polarization of the quantum vacuum by the gravitational field of the baryonic matter. Hence, for the first time, the quantum vacuum fluctuations, well established in quantum field theory but mainly neglected in astrophysics and cosmology, are related to the problem of dark matter."

The vorticity reaches a maximal value in the center of the vortex core.
We may imagine that the vorticity spreads far from the plane of a spiral galaxy in the transverse direction due to the involvement in the vortex activity of enormous amount of the virtual gravitational dipoles.
Such a direction is shown by dotted arrow in Fig.~\ref{fig=2}.
 It is instructive to recall here the Helmholtz theorems pointing to the certain properties of the vortex:
(i)~the strength of a vortex filament is constant along its length; (ii)~a vortex filament cannot end in a fluid medium; it must form a closed path (like a smoke ring), end at a boundary (solid or free surface), or go to infinity; (iii)~in the absence of rotational external forces, a fluid that is initially irrotational remains irrotational. 

One might guess that a substance called as
 the dark matter~\citep{Rubin2004, LaigleEtAl2014, PichonEtAl2014, PichonEtAl2014a} manifests itself through such vortex formations.
As follows from the  Helmholtz theorems they penetrate through the whole universe by creating a complex web of the
vortex filaments. 
These formations, thread-like formations, play a role in the centers of condensation of the barion matter along which the galaxies are formed.

\section{\label{sec5}Conclusion}

The equation of the relativistic hydrodynamics, describing motion of the perfect fluid, 
plus the continuity equation can be reduced to the relativistic quantum equation, the Klein-Gordon equation,
as soon as the quantum potential is found.
In non-relativistic limit these equations reduce to the Schr\"odinger equation.

The problem of finding the quantum potential 
goes back to Fick's  laws written down in the relativistic sector.
A main essence of these laws is that they describe currents induced by gradients of the pressures
arising in a special superfluid medium -  the physical vacuum.
So, the quantum potential describes influence through differences of the pressures
that arise between ensembles of virtual particles populating the vacuum.

The equation of the relativistic hydrodynamics when the viscosity term added gives the possibility of describing vortices arising in such a non-perfect fluid.
We proclaim in that regard that averaged in time the viscosity coefficient vanishes, but its dispersion is not zero.
Due to this trick, the vortex can live infinitely long and the vortex radius trembles.
The trembling is due to the exchange of the vortex energy with the zero-point vacuum fluctuations.
Note that, the vortex has non-zero radius of the core where the vortex speed tends to zero.

Consideration of the rotating motion of the spiral galaxy through a prism of the hydrodynamical equation loaded by the viscosity fluctuating about zero gives explanation for observed flattening of the orbital speeds on radii far exceeding the radius of the galactic core. Exchange of the energy of the rotating galactic matter with the zero-point fluctuations of the physical vacuum on ultra-low frequencies, that is, the exchanging by means of gravitational quanta, induces effect that looks as a breathing of the galaxy. Contribution of the ultra-low frequencies showing the distribution like the flicker noise law,  $1/f$, provides flattening the orbital speeds on the large radii. 

The zero-point vacuum fluctuations on the ulta-low frequencies can explain observed flattening of the orbital speeds of the spiral galaxies without attracting the idea of dark matter.

The Helmholtz theorem says that any vortex tube is continued in the space until it reaches a boundary of the space or the tube is closed on its own ends by forming a torus. It means that the vorticity in large-scale supports existence of the cosmic filaments. As observation shows, the cosmic filaments form a complicated web~\citep{LaigleEtAl2014, PichonEtAl2014, PichonEtAl2014a}. One may guess that this complicated web can be the result of turbulent mixing of the vortices. We know from the hydrodynamics, that the hydrodynamic equations contain a rich set of solutions. Among all of these, turbulent solutions represent a special class which is attracting attention by its variety of forms.

\begin{acknowledgments}

 The author thanks
 Mike Cavedon,  Pat Noland, and Denise Puglia
  for  valuable remarks and propositions.

\end{acknowledgments}


\end{document}